\documentclass[pra,aps,superscriptaddress,twocolumn,floats,10pt,floatfix,nobibnotes,accepted=2018-08-28]{quantumarticle}

\usepackage{amsmath}
\usepackage{amsfonts}
\usepackage{amssymb}
\usepackage{graphicx}
\usepackage{color}
\usepackage[usenames,dvipsnames,svgnames,table]{xcolor}
\usepackage[colorlinks=true]{hyperref}
\usepackage{amstext}
\usepackage{latexsym}
\usepackage{bbm}

\newcommand{\eref}[1]{Eq.~\eqref{#1}}
\newcommand{\fref}[1]{Fig.~\ref{#1}}

\newcommand{\aref}[1]{Appendix \ref{#1}}

\newcommand{\mbvd}{MBVD}

\newcommand{\fbar}{FBAR}

\begin{document}

\title{Electro-mechanical Casimir effect}
\author{Mikel Sanz}
\email{mikel.sanz@ehu.eus}
\affiliation{Department of Physical Chemistry, University of the Basque Country UPV/EHU, Apartado 644, E-48080 Bilbao, Spain}
\author{Witlef Wieczorek}
\affiliation{Department of Microtechnology and Nanoscience, Chalmers University of Technology, Kemiv\"agen 9, SE-41296 G\"oteborg, Sweden}
\author{Simon Gr\"oblacher}
\affiliation{Kavli Institute of Nanoscience, Delft University of Technology, Lorentzweg 1, 2628CJ Delft, The Netherlands}
\author{Enrique Solano}
\affiliation{Department of Physical Chemistry, University of the Basque Country UPV/EHU, Apartado 644, E-48080 Bilbao, Spain}
\affiliation{IKERBASQUE, Basque Foundation for Science, Maria Diaz de Haro 3, E-48013 Bilbao, Spain}
\affiliation{Department of Physics, Shanghai University, 200444 Shanghai, China}

\begin{abstract}
The dynamical Casimir effect is an intriguing phenomenon in which photons are generated from vacuum due to a non-adiabatic change in some boundary conditions. In particular, it connects the motion of an accelerated mechanical mirror to the generation of photons. While pioneering experiments demonstrating this effect exist, a conclusive measurement involving a mechanical generation is still missing. We show that a hybrid system consisting of a piezoelectric mechanical resonator coupled to a superconducting cavity may allow to electro-mechanically generate measurable photons from vacuum, intrinsically associated to the dynamical Casimir effect. Such an experiment may be achieved with current technology, based on film bulk acoustic resonators directly coupled to a superconducting cavity. Our results predict a measurable photon generation rate, which can be further increased through additional improvements such as using superconducting metamaterials.
\end{abstract}

\maketitle

\section{Introduction}\label{sec:introd}
Quantum mechanics predicts that virtual particles can emerge from vacuum. This phenomenon, known as quantum fluctuations, is a cornerstone to explaining key effects in nature, ranging from the anomalous magnetic moment of the electron~\cite{QFTBook}, to the enhancement in quantum transport phenomena~\cite{dV08}, and the Lamb shift of atomic spectra~\cite{Lamb1947}. Another paramount example is the Casimir effect, which results from a force between two separated conducting plates ~\cite{Casimir1948,Casimir1997,Casimir1998,lamoreaux_progress_2011}. 
G. T.\ Moore~\cite{DCasimir} suggested the existence of a kinetic analogue to this phenomenon, known as the dynamical Casimir effect (DCE). This effect, based on the fact that a moving mirror modifies the mode structure of the electromagnetic vacuum dynamically, is the result of a mismatch of vacuum modes in time. Indeed, if the velocity of a mirror is much smaller than the speed of light, no excitations emerge out of the vacuum, since the electromagnetic modes adiabatically adapt to the changes. However, if the mirror experiences relativistic motion, these changes occur non-adiabatically. Hence, the vacuum state of the electromagnetic field can be excited by taking energy from the moving mirror, resulting in the generation of real photons. It has been suggested that, due to the large stress generated on any macroscopic material moving with relativistic speed, the observation of the DCE might actually be unrealistic \cite{braggio_novel_2005,dodonov_current_2010}. This challenge inspired various alternative proposals to observe the DCE \cite{yablonovitch_accelerating_1989,LJR96,dodonov_generation_1996,ji_production_1997,uhlmann_resonant_2004,crocce_model_2004,braggio_novel_2005,kim_detectability_2006,gunter_sub-cycle_2009,de_liberato_extracavity_2009,johansson_dynamical_2009,JJWN10,nation_ultrastrong_2016}, e.g., by employing nanomechanical resonators, modulating two-level systems inside a cavity or modulating the electric boundary conditions of a superconducting cavity; for an overview see, e.g.,~\cite{dodonov_current_2010,dalvit_fluctuations_2011,NJBN12}.

Only recently, DCE radiation has been experimentally observed \cite{CMWilson2011} in superconducting circuits by moving the electric boundary condition of a superconducting cavity \cite{johansson_dynamical_2009,JJWN10} instead of the center of mass of a mirror. In a similar experiment, DCE photons were created by modulating the effective speed of light~\cite{Hakonen2013} making use of a superconducting metamaterial. Additionally, relevant applications of these experiments, such as a robust generation of multi-partite quantum correlations~\cite{johansson_dynamical_2009,JJWN10,GAL2010,johansson_nonclassical_2013,FSLRJDS14,RFERSS16}, were proposed, and recently observed~\cite{Schneider18}, thus, highlighting the potential of this effect beyond its fundamental interest. However, no experiment to date has succeeded in generating photons out of vacuum using a moving mechanical object in the spirit of the original DCE proposal. 

\begin{figure}[thbp] 
  \centering
      \includegraphics[width=0.45\textwidth]{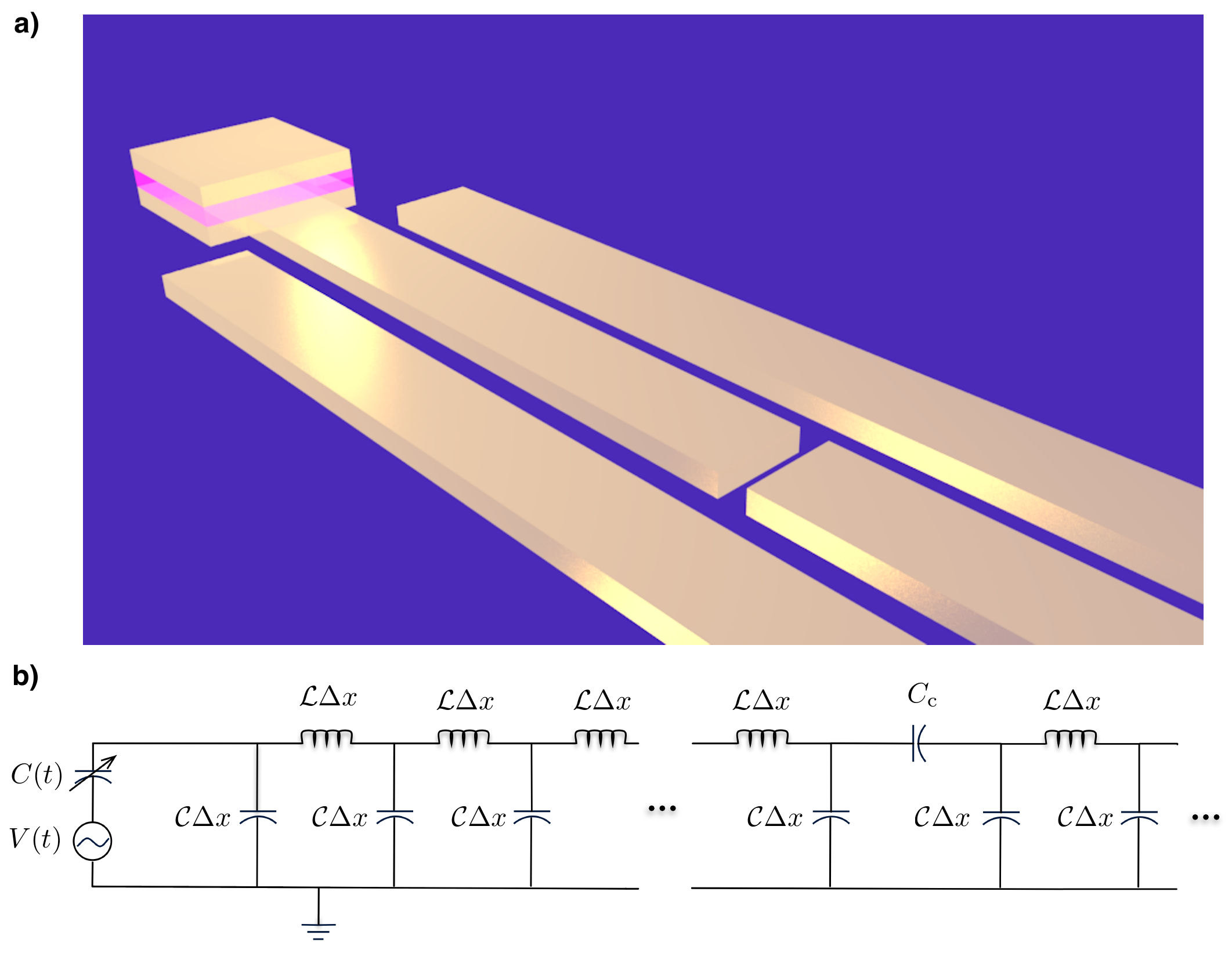}
      \caption{Scheme of the coupling between an FBAR and a superconducting cavity. (a) Artistic view representing an FBAR (top left), i.e., a resonator composed of two superconducting layers sandwiching a piezoelectric material. The FBAR terminates one side of a superconducting cavity that is capacitively coupled to a semi-infinite transmission line (lower right). (b) Equivalent circuit of the proposed implementation, which is composed of the mechanical resonator (time-dependent capacitance $C(t)$ and voltage drive $V(t)$) and the superconducting cavity (with $\mathcal{C}$ and $\mathcal{L}$ capacitance and inductance per unit length $\Delta x$). We have employed the modified van Dyke-Butterworth model to provide the equivalent lumped-element circuit representation of the FBAR and model the cavity as an infinite set of LC circuits which will be afterwards capacitively coupled via $C_c$ to a semi-infinite waveguide.}\label{fig:coupling}
\end{figure}

In this letter, we propose an experiment consisting of a nanomechanical resonator that is directly coupled to a high impedance superconducting cavity (see Fig.~\ref{fig:coupling}) in order to create a measurable rate of electro-mechanically generated DCE photons. Our proposal develops the idea to employ film bulk acoustic resonators (FBAR) for DCE photon creation from Ref.~\cite{kim_detectability_2006} further by combining it with the technology of superconducting circuits~\cite{johansson_dynamical_2009,JJWN10,OHAB10,CMWilson2011}. Our calculations make use of recent advances in mechanical oscillators based on FBAR technology using thin films of Al-AlN-Al~\cite{OHAB10}. The presence of a superconducting cavity allows for a resonant enhancement of the photon rate which surpasses the radiation from a single mirror by several orders of magnitude~\cite{LJR96,dodonov_generation_1996,ji_production_1997}. We study the minimally required conditions to observe a stable flux of photons resulting from the electro-mechanically amplified quantum vacuum, concluding that such a measurement is feasible with current technology. Finally, we propose technical improvements to further enhance the mechanical photon production. Our work also paves the way to experimentally test other fundamental relativistic effects with mechanical resonators, such as the Unruh effect or the Hawking radiation~\cite{NJBN12} 

\section{Proposed implementation}
The proposed setup consists of a piezoelectric FBAR resonator with resonance frequency  $\Omega/2\pi = 4.20$~GHz (e.g.~made from Al-AlN-Al) and a superconducting cavity (e.g.~made from Al) with length $d = 33$~mm, which is capacitively coupled to a semi-infinite transmission line for the read-out, as shown in Fig.~\ref{fig:coupling}, all assumed to be operating at a temperature of $10\,$mK. To match the resonance condition in an actual experimental implementation, we note that the superconducting cavity can be made tunable in frequency by incorporating a flux-biased SQUID \cite{CMWilson2011,Sandberg2009}. We have chosen values that closely follow the parameters from Ref.~\cite{OHAB10}, for details see \aref{app:params}. An AC voltage is applied to the superconducting plates of the mechanical resonator. This voltage drive has two effects: (i) the piezoelectric material of the \fbar{} converts the AC voltage into mechanical contraction and expansion, which essentially leads to a change of the length of the superconducting cavity; (ii) the voltage drive changes the potential of the cavity's boundary. Both effects result in the production of DCE photons.

\subsection{FBAR modeling}
We make use of the modified van Dyke-Butterworth model to construct the equivalent lumped-element circuit representation of the FBAR~\cite{LBWR00,N05} (for details see \aref{app:model}). This circuit contains two parallel branches. The first one consists of a mechanical capacitance $C_m$, mechanical inductance $L_m$ and a resistor $R_m$ modeling the mechanical dissipation connected in series, while the second one contains a resistor $R_0$ taking into account the dielectric loss and, crucially, a geometrical capacitance $C(t)$, which slightly changes in time when an AC voltage is applied to the plates of the FBAR. Composing the impedances for both branches, $Z_m$ and $Z_0$, it is straightforward to prove that the total impedance $Z \approx Z_0$, which, to first order, can be reduced to $C(t)$ (see \aref{app:model}).

Let us estimate the change in the capacitance $C(t)$ when an AC voltage $V(t) = V_{\text{pp}} \cos (\omega t + \phi)$ is applied to the electrodes. An AC voltage with an amplitude $V_{\mathrm{pp}}$ at the resonance frequency of the mechanical resonator $\omega=\Omega$ results in a change of the inter-plate distance $\Delta x (V_{\mathrm{pp}}) \approx1.7\cdot V_{\mathrm{pp}}$ nm/V (see \aref{app:coupling}). Crucial here is that the piezoelectric effect is enhanced by the mechanical quality factor of the FBAR, assumed to be $Q=300$ \cite{OHAB10}. Applying a voltage of $V_{\mathrm{pp}}=500\mu$V thus results in $\Delta x=8.5\cdot 10^{-13}$m, which is more than five orders of magnitude smaller than the thickness $t_{\text{AlN}}=3.5\cdot 10^{-7}$m of the piezoelectric layer. Thus, we model the resulting mechanical contraction and expansion as harmonic. The time-varying capacitance is then given as $C(t) = \varepsilon_{\text{AlN}}A/[t_{\text{AlN}}+ \Delta x \cos (\Omega t)]$, with $A$ as area of the \fbar{} and $\varepsilon_{\text{AlN}}$ as dielectric constant of AlN. We expand $C(t)$ for $\Delta x\ll t_{\text{AlN}}$ and finally obtain
\begin{equation}
C(t) \approx C_0 + \Delta C \cos (\Omega t)
\end{equation}
with $C_0=\varepsilon_{\text{AlN} }A/t_{\text{AlN}}$ and  $\Delta C \approx C_0 \Delta x/t_{\text{AlN}}$.\\

\subsection{Lumped-element circuit model}
The Lagrangian describing the circuit of the FBAR connected to an open transmission line (without the coupling capacitor $C_c$ shown in Fig.~\ref{fig:coupling}) can be written as

\begin{eqnarray} \label{lagrangianFBAR}
L &=& \sum_{i=0}^\infty \left [ \frac{\delta x \mathcal{C}}{2} \dot{\Phi}_{i+1}^2 - \frac{1}{2 \delta x \mathcal{L}}  (\Phi_{i+1} - \Phi_{i})^2 \right ] \nonumber \\
&+& \frac{1}{2} C(t) (\dot{\Phi}_0 - \dot{\Phi}_v)^2 + \frac{1}{2} C_g \dot{\Phi}_0^2.
\end{eqnarray}
Here, $\mathcal{C}$ and $\mathcal{L}$ are the densities of capacitance and inductance of a transmission line per unit length $\delta x$, respectively, $C_g$ is the capacitive coupling of the FBAR to ground, which we will discard in our analysis as it is much smaller than any other quantity involved, $\Phi_i$ is the $i$-th node flux and $\dot{\Phi}_v=V$ is the voltage of the AC source. In the continuous limit, the equation of motion corresponding to $i=0$ is given by

\begin{equation}\label{eqmotion}
C(t) \ddot{\Phi}(0,t) + \dot{C}(t)\dot{\Phi}(0,t)  -\frac{1}{\mathcal{L}} \frac{\partial \Phi (x,t)}{\partial x}\Big |_{x=0} = F(t),
\end{equation}
with $F(t) = \frac{d}{dt} (\theta(t) C(t) V(t))$ being the electro-mechanical source term, and $\theta(t)$ is the Heaviside step function. In order to solve it, we follow a similar approach to Ref.~\cite{johansson_dynamical_2009,JJWN10} and expand the field in the Fourier components

\begin{align*}
\Phi(x,t) = \sqrt{\frac{\hbar Z_0}{4 \pi}} & \int_{0}^{\infty} d\omega \frac{1}{\sqrt{\omega}} \left [ a_\text{in}(\omega)e^{-i (\omega t - k_{\omega} x)}  \right. \\ 
& \left.  + a_\text{out}(\omega) e^{-i (\omega t + k_{\omega} x)} + H.c. \right ],
\end{align*}

with $Z_0 \approx 55$~$\Omega$. Equation~\eqref{eqmotion} can now be written as

\begin{align*}
\int_{0}^{\infty}  d\omega & \sqrt{\frac{\hbar Z_0}{4 \pi |\omega|}}  \Bigg[  &&\\
&+\left (-\omega^2 C(t) - i \omega \dot{C}(t)-\frac{i k_{\omega}}{\mathcal{L}}\right) a_\text{in}(\omega) e^{-i\omega t}  \\
& + \left (-\omega^2 C(t) - i \omega \dot{C}(t) + \frac{i k_\omega}{\mathcal{L}} \right ) a_\text{out}(\omega) e^{-i\omega t} \\
 &  + H.c. \Bigg] = F(t).
\end{align*}

By integrating over $\int_{-\infty}^{\infty} dt \sqrt{|\omega'|} e^{i \omega' t}$, we can see that in the case of a static capacitor (i.e.~$C(t)=C_0$), it behaves as a mirror placed at $x = - L_\text{eff} \approx \frac{C_0}{\mathcal{C}}$, such that the effective length of the resonator shown in Fig.~\ref{fig:coupling}, which will be introduced below, is $d_\text{eff} = d + L_\text{eff}$.

Resonances emerge due to the inelastic interaction of the photons with the oscillating mirror, and the incoming modes $a_\text{in}(\omega)$ with frequency $\omega$ are scattered elastically as $a_\text{out}(\omega)$ and inelastically as $a_\text{out}(\omega+\Omega)$, $a_\text{out}(\omega+ 2\Omega)$, ... . Keeping only the first inelastic resonances, as higher resonances have higher orders in $\Delta C$, and assuming that $0 < \omega \leq \Omega$, which is natural since the condition for the rotating wave approximation holds around $\Omega \approx 2 \omega'$, the output mode is given by 

\begin{align}
a_\text{out}(\omega, L_\text{eff}) & = h(\omega, \Omega) + a_\text{in}(\omega, L_\text{eff}) \nonumber \\
& + S(\omega,\Omega+\omega) a_\text{in}(\Omega+\omega, L_\text{eff})  \nonumber \\
&  + S(\omega,\Omega-\omega) a_\text{in}^{\dagger}(\Omega-\omega,L_\text{eff}).
\end{align}

Here, the operators $a_\text{in}$ and $a_\text{out}$ are defined in the displaced position $x=L_\text{eff}$, $h(\omega,\Omega)$ is the identity in operator space, as we have treated the source classically, and

\begin{align}
S(\omega',\omega'') & = -i \Delta C Z_0 \sqrt{|\omega'| |\omega''|} \theta(\omega') \theta(\omega''), \label{eq:S}\\
h(\omega,\Omega) & =  -i \sqrt{\frac{4 \pi Z_0}{\hbar \omega}} \mathcal{F} (\omega, \Omega), \label{eq:h}
\end{align}
where $\mathcal{F} (\omega, \Omega)= (2\pi)^{-1/2} \int_{-\infty}^{\infty} dt F(t) e^{i \omega t}$ is the Fourier transform of $F(t)$. 

The aforementioned result describes an oscillatory mirror coupled to a one-dimensional open transmission line. The photon production due to the change in the boundary conditions can be dramatically increased by introducing a cavity with a well chosen resonance condition, as shown for example in Ref.~\cite{LJR96,dodonov_generation_1996,ji_production_1997}. In order to realize this, we will introduce a capacitor at a distance $d$ of the FBAR, as depicted in Fig.~\ref{fig:coupling}. We redefine the origin of the coordinates in the coupling capacitance between the cavity and the line and hence the harmonic mirror is placed at $x= d_\text{eff} = d + L_\text{eff}$, with $d$ the length of the cavity for a static mirror. Therefore, the natural frequency of the resonator is $\omega_0/2\pi = v/ d_\text{eff}$ with $v$ the speed of light in the superconducting material. The input-output relations connecting the fields inside the cavity $\{ a_\text{in}, a_\text{out} \}$ to fields in the transmission line $\{b_\text{in}, b_\text{out} \}$ are given by

\begin{equation} \label{fields}
\begin{pmatrix}
a_\text{in}(\omega,0) \\
a_\text{out}(\omega,0)
\end{pmatrix}
= 
\begin{pmatrix}
\bar{\alpha}_\omega & \beta_\omega \\
\bar{\beta}_\omega & \alpha_\omega
\end{pmatrix} \,
\begin{pmatrix}
b_\text{in}(\omega,0) \\
b_\text{out}(\omega,0)
\end{pmatrix},
\end{equation}
with $\alpha_\omega = 1+i \omega_c/2\omega$, $\beta_\omega = i\omega_c/2\omega$ and the coupling rate $\omega_c = 1/C_c Z_0$. Notice that $\omega_c$ is a measure of how large the coupling of the resonator to the line is. Indeed, for $\omega_c \gg \omega$ it is completely decoupled, while $\omega_c = 0$ means that there is no cavity. Equation \eqref{fields} holds for $x=0$ \cite{johansson_dynamical_2009,JJWN10}, but as we want to connect the fields in the line with the reflected fields on the oscillating mirror, we must displace the cavity fields to $x=d_\text{eff}$ by using the matrix $\text{diag} (e^{i k_{\omega}d_\text{eff}},e^{-i k_{\omega}d_\text{eff}})$. In order to compute the reflection coefficient $R_\text{res}(\omega)$ of the cavity, we can consider that the inelastic scattering process is absent and that only the reflection-transmission process remains, so $a_\text{in}(\omega,d_\text{eff}) = a_\text{out}(\omega,d_\text{eff})$. Under this condition, we get

\begin{equation}
R_\text{res}(\omega) = \frac{1+(1+\frac{2 i \omega}{\omega_c})e^{2 i k_{\omega} d_\text{eff}}}{(1- \frac{2 i \omega}{\omega_c})+e^{2 i k_{\omega} d_\text{eff}}}.
\end{equation} 

Similarly, we also want to study the mode structure of the resonator. This information is encoded in the function $A_\text{res}$, defined as $a_\text{out}(\omega,d_\text{eff}) = A_\text{res}(\omega)b_\text{in}(\omega,0)$, as it contains the response of the resonator to any input signal coming from the line,

\begin{equation}
\label{Ares}
A_\text{res} (\omega) = \frac{(\frac{2 i \omega}{\omega_c})e^{i k_{\omega} d_\text{eff}}}{(1- \frac{2 i \omega}{\omega_c})+e^{-2 i k_{\omega} d_\text{eff}}}.
\end{equation} 

The resonance frequencies can be deduced from the denominator of Eq.~\eqref{Ares} and, as shown in Ref.~\cite{JJWN10}, they are approximately the solutions of the transcendental equation $\tan (2 \pi \omega / \omega_0) = \omega_c / \omega$. The outgoing mode in the open transmission line in the presence of a driven FBAR, keeping only the first order of $S(\omega',\omega'')$, reads

\begin{align}
b_\text{out}(\omega) & = h_\text{res}(\omega,\Omega) + R_\text{res}(\omega) b_\text{in}(\omega) \nonumber \\
& + S_1^\text{res}(\omega,\Omega+\omega) b_\text{in}(\Omega+\omega) \nonumber \\
& + \bar{S}_{2}^\text{res}(\omega,\Omega-\omega) b_\text{in}^{\dagger}(\Omega-\omega),
\end{align}
with
\begin{align*}
h_\text{res}(\omega,\Omega) & = h(\omega,\Omega) \left[(1- \frac{2 i \omega}{\omega_c})+e^{-2 i k_{\omega} d_\text{eff}}\right]^{-1}\\
S_1^\text{res}(\omega',\omega'') & = S(\omega',\omega'') A_\text{res}(|\omega'|) A_\text{res}(|\omega''|) \\
S_2^\text{res}(\omega',\omega'') & = S(\omega',\omega'') \bar{A}_\text{res}(|\omega'|) A_\text{res}(|\omega''|).
\end{align*}

\section{Electro-mechanical DCE photon rate}
For an initial thermal state, the mean photon number $n_\text{out} (\omega) = \langle b_\text{out}^{\dagger}(\omega) b_\text{out} (\omega) \rangle_{T}$ is given by
\begin{eqnarray}
\label{eq:Nout}
\nonumber n_\text{out}(\omega) & = & \left |R_\text{res}(\omega)\right |^2 n_\text{in}(\omega)\nonumber \\
& + & \left | S_1^\text{res}(\omega, \Omega+\omega)\right |^2 n_\text{in}(\Omega + \omega) \nonumber \\ 
& + & \left | S_2^\text{res}(\omega,\Omega-\omega)\right |^2 \left[1+n_\text{in}(\Omega - \omega)\right] \nonumber\\
& + & |h_\text{res}(\omega,\Omega)|^2,
\end{eqnarray}

\begin{figure}[t] 
  \centering
      \includegraphics[width=0.40\textwidth]{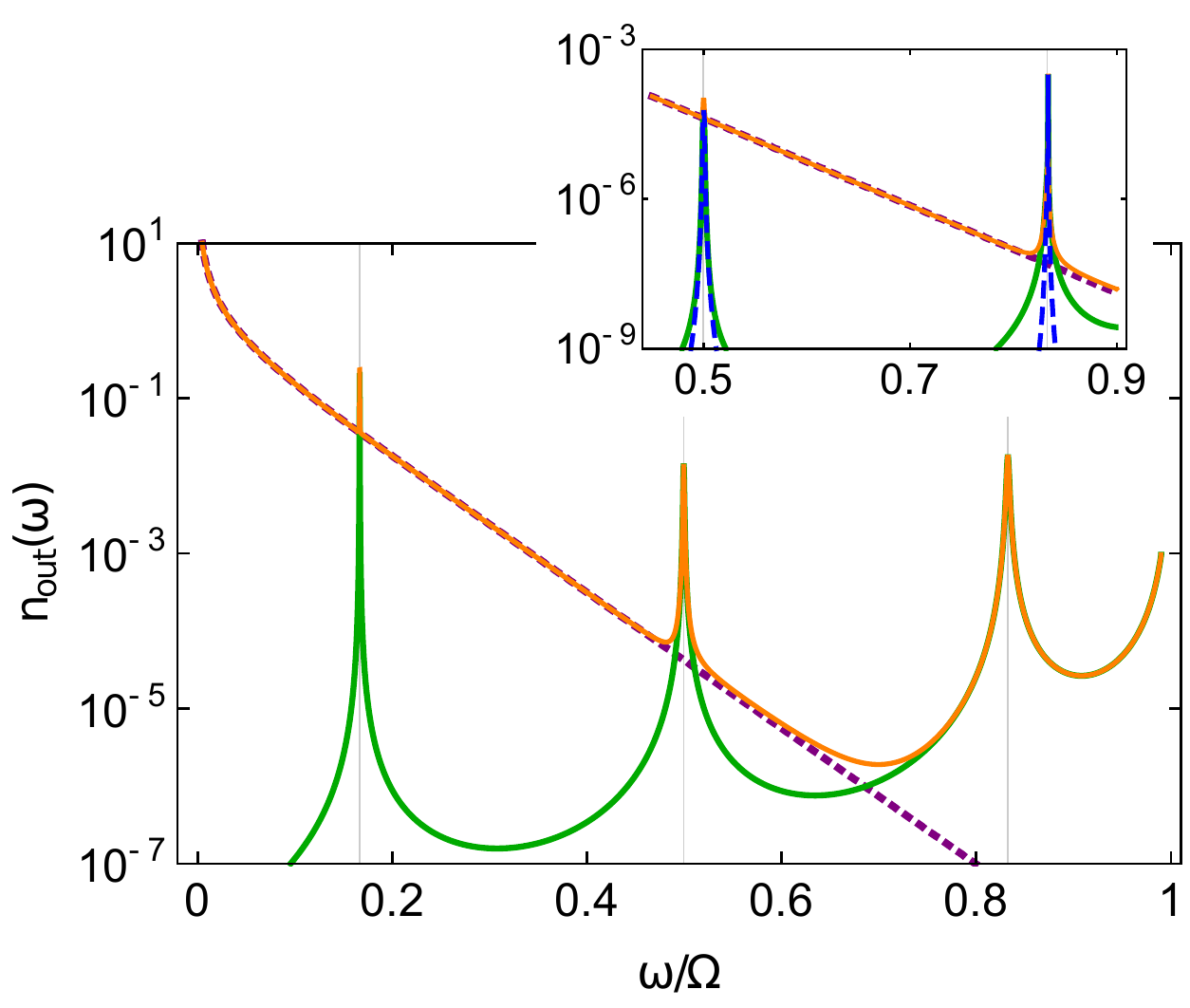}
      \caption{Electro-mechanical DCE photon production. We consider the generated photons $n_{out}(\omega)$ in Hz per unit bandwidth versus observation frequency $\omega$ in units of the mechanical frequency $\Omega$. Main panel (using $V_{\mathrm{pp}}=500\mu$V, $Q=300$): The orange (light gray) line depicts the total number of photons in the open transmission line that originate from the DCE effect and thermal radiation. The green (dark gray) line shows the photons only generated via the electro-mechanical DCE effect, whereas the purple (dotted) line shows the thermal contribution. The vertical gray lines depict the resonances of the superconducting cavity. The inset shows an enhanced mechanical DCE photon production assuming an \fbar{} with a higher mechanical quality factor ( $Q=3\cdot 10^6$ and using a driving voltage of $V_{\mathrm{pp}}=5\mu$V). The additional blue (dashed) line shows the DCE photons only created through the mechanical motion of the FBAR.}\label{fig:DCEphotons}
\end{figure}

where $n_\text{in}(\omega) = \left [\exp (\hbar \omega/k_B T) -1 \right ]^{-1}$ is the thermal photon occupation at temperature $T$. The photon production of electro-mechanical Casimir origin is given by the term $| S_2^\text{res}(\omega,\Omega-\omega) |^2 + |h_\text{res}(\omega,\Omega)|^2$. The total mean photon number $n_\text{out}(\omega)$ is plotted in Fig.~\ref{fig:DCEphotons}. For $\omega/\Omega=1/2$ we see that the Casimir photon production is of the order of $10^{-2}$ photons/s/unit bandwidth for a driving voltage of $V_{\text{pp}}= 500~\mu$V. This is comparable to the photon production predicted by \cite{johansson_dynamical_2009,JJWN10} and should be easily measurable with current technology~\cite{Metal10,DCMZDMGS14}. The DCE part of the total photon number is well above the thermal photon noise floor.

The non-adiabatic change in the boundary conditions is of electro-mechanical nature, as the voltage driving the \fbar{} also produces a change in the potential of the boundary, in addition to the mechanical movement of the plate. Even though it can not be measured, one may wonder about the ratio of the photon production which is purely due to the mechanical movement. In order to calculate electrically generated photons, we consider the situation in which the piezoelectric material is removed, i.e., $\Delta C = 0$, while the voltage source is still connected. By subtracting this from the total number of generated photons with the same parameters as before, we estimate the purely mechanical photon production to be $5\times 10^{-9}$ photons/s/unit bandwidth. Therefore, only about one in every million generated photons can be ascribed to mechanical motion alone. However, in our situation it is not possible to experimentally distinguish the photons generated by changes in the electrical boundary conditions from the ones generated by changes in the mechanical boundary conditions. Consequently, we can only talk about electro-mechanically produced photons. 

Improving the quality factor of the \fbar{} will drastically change the picture. Assuming a quality factor of $3\cdot10^6$, i.e.~4 orders of magnitude larger than in Ref.~\cite{OHAB10}, and a smaller driving voltage of $V_{\text{pp}}=5\,\mu$V results in $\Delta x=8.5\cdot10^{-11}$~m. This high-Q regime is in principle already experimentally accessible with other materials, for example with GHz mechanical oscillators made of silicon~\cite{MEEN14}. The higher mechanical quality factor and the lower driving voltage benefit the ratio of DCE photons generated via mechanical compared to electrical origin, such that both can be of equal magnitude, see inset of \fref{fig:DCEphotons}.

\section{Discussion and Conclusion}
Further improvements could be obtained by using a material with higher piezoelectric coefficient, higher mechanical quality factor and larger dielectric constant for the \fbar{}. The most promising approach is the use of SQUID-based metamaterials for the resonator. Indeed, recent experiments with metamaterials based on long arrays of Josephson junctions have demonstrated \cite{MPKMD12,WKDFMNBHG15} an increase in the effective impedance $Z_0$ by two orders of magnitude to $Z_0 \approx 10^4$~Ohm. This was achieved while keeping the total capacitance of the resonator approximately constant, which means that the speed of light in the metamaterial $v = (\mathcal{C} Z_0)^{-1}$ is reduced by two orders of magnitude. Such a device would lead to an enhancement in photon production by four orders of magnitude, from which we can again roughly estimate the ratio of the mechanically produced photons with respect to the electrically generated ones, which would improve by two orders of magnitude. This results from the former scaling with $Z_0$, as shown in Eq.~\eqref{eq:S}, while the latter scaling as $Z_0^{1/2}$, as shown in Eq.~\eqref{eq:h}. However, using superconducting metamaterials could lead to other unwanted effects such as the emergence of a finite-size lattice structure or the impedance mismatch between the resonator and the transmission line. Incorporating these properly into our proposal would require more detailed studies, which we leave as a future direction.

Finally, we would like to highlight that there are other traces of the Casimir effect related to quantum correlation functions. For instance, techniques developed in recent accurate experiments to measure two-time correlation functions in propagating quantum microwaves~\cite{DCFZFMSDMGS15,Fetal16,Fetal17} can be used to measure $g^{(2)}(\tau)$. Indeed, the photons which leak out of the cavity will show a decaying behavior of these auto-correlations, however the formalism developed in this letter is not suitable to calculate such effects, as we treat both the capacitor and the voltage source as classical elements.

In summary, we have proposed an experiment consisting of a mechanical resonator based on a FBAR directly coupled to a superconducting cavity, which generates a resonance enhancement of the electro-mechanically generated dynamical Casimir radiation. We calculate the stable flux of photons proceeding from an electro-mechanically amplified quantum vacuum, demonstrating that measuring an effect is within reach of current technology. We also propose a heuristic method to estimate the fraction which can be ascribed to the non-adiabatic change in the mechanical boundary conditions. Further improvements can either be realized by using Josephson metamaterials or a larger mechanical quality factor to enhance both the photon production and the ratio of the purely mechanically generated photons. Our work also  paves the way towards experimentally tests of other fundamental relativistic effects, such as the Unruh effect or the Hawking radiation, by properly modifying the proposals in Ref.~\cite{NJBN12} employing mechanical resonators.

\begin{acknowledgments}
We would like to thank Adrian Parra, Simone Felicetti, Philipp Schmidt, Hans Huebl, Nicola Roch, and Gary Steele for fruitful discussions and useful insights. M.S. and E.S. are grateful for funding through the Spanish MINECO/FEDER FIS2015-69983-P and Basque Government IT986-16. S.G. acknowledges financial support from Foundation for Fundamental Research on Matter (FOM) Projectruimte grants (15PR3210, 16PR1054), the European Research Council (ERC StG Strong-Q), and the Netherlands Organisation for Scientific Research (NWO/OCW), as part of the Frontiers of Nanoscience program, as well as through a Vidi grant (016.159.369). W.W. acknowledges financial support by Chalmers Excellence Initiative Nano.
\end{acknowledgments}

\appendix 

\section{Piezo-electromechanical coupling}\label{app:coupling}

We base our proposal on an \fbar{} device of thickness $t$ and square surface area $A=b^2$ with volume $V=A\cdot t$ and define the z-direction (subscript $3$) along the thickness of the \fbar{}. A voltage is applied to the plates of the \fbar{}, resulting in an electric field along the z-direction that is assumed to be homogeneous. Due to piezoelectric coupling this voltage results in mechanical strain $\epsilon$:
\begin{equation}\label{eq:pm}
\epsilon_{3j}=-d_{3j}E_3=-d_{3j}\frac{V_3}{t},
\end{equation}
where $d_{ij}$ is the piezoelectric coupling coefficient. This mechanical strain has a geometric and stress-related effect, which are discussed in the following.

\subsection{Geometric effect}

Strain leads to contraction and expansion of the mechanical device and, thus, its dimensions change. For the $z$ direction, one obtains:
	\begin{equation}
		\Delta z=\epsilon_{33}\cdot t=-d_{33}V_3.
	\end{equation}
This dimensional change shifts the resonance frequency of the \fbar{} device. Using $\Omega=2\pi v/2t$ ($v$ velocity of sound) as approximate formula for the resonance frequency of an \fbar{}, one gets $\Delta \Omega/\Omega_0=-\Delta t/t\approx -\epsilon_{33}$. Using \eqref{eq:pm}, one arrives at
\begin{equation}
    \frac{\Delta \Omega}{\Omega_0}=d_{33}\frac{V_3}{t}.
\end{equation}

\subsection{Stress-related effect}

Mechanical strain $\epsilon$ results in mechanical stress $\sigma$ via the relation $\sigma=E\epsilon$ ($E$ is Young's modulus). Mechanical stress over an area $A$ results in a force $F$ acting on the crystal surface $F=\int_A\sigma dA$. 

This force can change the resonance frequency of the device, if the crystal structure could not relax along that direction. For example, this would be the case for a doubly-clamped mechanical beam. However, the \fbar{} has no fixed/clamped boundaries along the z-direction and, thus, can relax. Hence, we neglect any shift in resonance frequency of the \fbar{}.

The force $F$ can nevertheless result in driving the mechanical amplitude of the resonator as it acts as an additional source term in the dynamic equation. The solution of a driven, damped harmonic oscillator in Fourier space is
\begin{equation}
	\tilde{x}(\omega)=\chi(\omega)\cdot \tilde{F}(\omega)/m,
\end{equation}
with the mechanical susceptibility $\chi(\omega)=(\Omega^2-\omega^2-i\gamma\omega)^{-1}$ and the mass $m$ of the resonator. A sinusoidal driving voltage $V_3(t)=V_{\text{pp}}\cos{(\omega t+\phi)}$ results in a force 
\begin{eqnarray}
	F_3(t)&=& \int_A \sigma_3(t)dA=Ed_{33}\frac{V_3(t)}{t}\int_AdA \nonumber \\
	&=& Ed_{33}\frac{V}{t^2}V_3(t).
\end{eqnarray}
Evaluating the mechanical response $\tilde{x}(\omega)$ on resonance $\omega=\Omega$ yields
\begin{equation}\label{amplitude}
    |\tilde{x}(\Omega)|=\frac{Q}{\Omega^2}\frac{E}{\rho t}d_{33}\frac{V_{\text{pp}}}{t}.
\end{equation}
This formula can be rewritten by using the \fbar{s} resonance frequency $\Omega=2\pi v/2t$ (with velocity of sound $v=\sqrt{K/\rho}$, bulk modulus $K=E/(3(1-2\nu))$, Poisson's ratio $\nu$, density $\rho$) to
\begin{eqnarray}
	\nonumber |\tilde{x}(\Omega)|&=&\frac{Q}{\pi^2}\left(3(1-2\nu)\right)d_{33}V_{\text{pp}}\\
	&=&\frac{Q}{\pi^2}\left(3(1-2\nu)\right)\Delta z.
\end{eqnarray}
This means that the driven response is by a factor of $\sim Q/\pi^2$ larger than the geometric response $\Delta z$ alone.

\section{Proposed implementation and modeling}\label{app:model}

\subsection{FBAR}

We base our proposal on existing technology and employ experimental parameters closely following Ref.~\cite{OHAB10}. The mechanical resonator is a piezoelectric FBAR, made up of a heterostructure from Al-AlN-Al, whereby the Al is used to contact the piezoelectric AlN. A specific device could have a total thickness of approximately $1000$~nm, consisting of a $300$~nm SiO$_2$ layer, two $150$~nm Al electrodes surrounding a $350$~nm AlN film, with an average speed of sound of $v_s = 9100$~m/s. The device could be fabricated on a high-resistivity silicon-on-insulator wafer. We drive the \fbar{} with an AC voltage source $V(t)= V_{\text{pp}} \cos (\omega t + \phi)$ with $\omega=\Omega$, i.e., a driving frequency equal to the fundamental mechanical frequency $\Omega$, and a phase difference $\phi = \frac{\pi}{2}$. 

\subsection{Modified Butterworth-Van Dyke circuit}\label{sec:MBVD}

The Modified Butterworth-Van Dyke circuit (\mbvd{}) \cite{LBWR00} enables to extract the necessary parameters for modeling a mechanical resonator as an equivalent electrical circuit. This has been, e.g., used in Ref.~\cite{OHAB10} for calculating the response of the \fbar{} made from Al/AlN/Al. Note that the \mbvd{} is an approximation of the Generalized Butterworth-Van Dyke circuit \cite{jin_generalised_2011} for low electro-mechanical coupling $k_t^2$.

The simplest model for a \fbar{} is derived from approximating the 3-port Mason model to a 4 element circuit consisting of a plate capacitance $C_0$ in parallel with a series $R_m-L_m-C_m$ circuit. In our case, the capacitance $C_0\equiv C(t)$ actually changes in time when an AC driving is applied, due to the change in the geometric structure. Indeed, mechanical resonators consisting of a piezoelectric material sandwiched between superconducting layers suffer from a geometrical change in their structure due to the piezoelectric effect when a voltage is applied. Therefore, the value of the electric elements describing the electric response of the resonator depends on the applied voltage (note that our voltage source is a function of time). For the usual applications of FBARs and other mechanical oscillators, this dependence is negligible, since it is a tiny correction of the mean value. However, this dependence has already been studied, for instance, in Ref.~\cite{Lee16}, and references thereof. The aforementioned 4-element circuit has a series resonance $\omega_s$ (given by $L_m,C_m$ as $\omega_s=1/\sqrt{L_mC_m}$) and a parallel resonance $\omega_p$ (set by $C_0$ in series with $L_m,C_m$ as $(\omega_p/\omega_s)^2=1+1/r$). The capacitance ratio $r$ is defined as $r=C_0/C_m$. The electro-acoustic coupling constant $k_t^2$ can be derived to be 

\begin{equation}
    (k_t)^2=\frac{\pi^2}{4}\frac{\omega_s}{\omega_p}\frac{\omega_p-\omega_s}{\omega_p}=\frac{\pi^2}{8}\frac{1}{r}\left(1-\frac{1}{r}\right)
\end{equation}

\begin{figure}[thbp]
  \centering
      \includegraphics[width=0.47
     \textwidth]{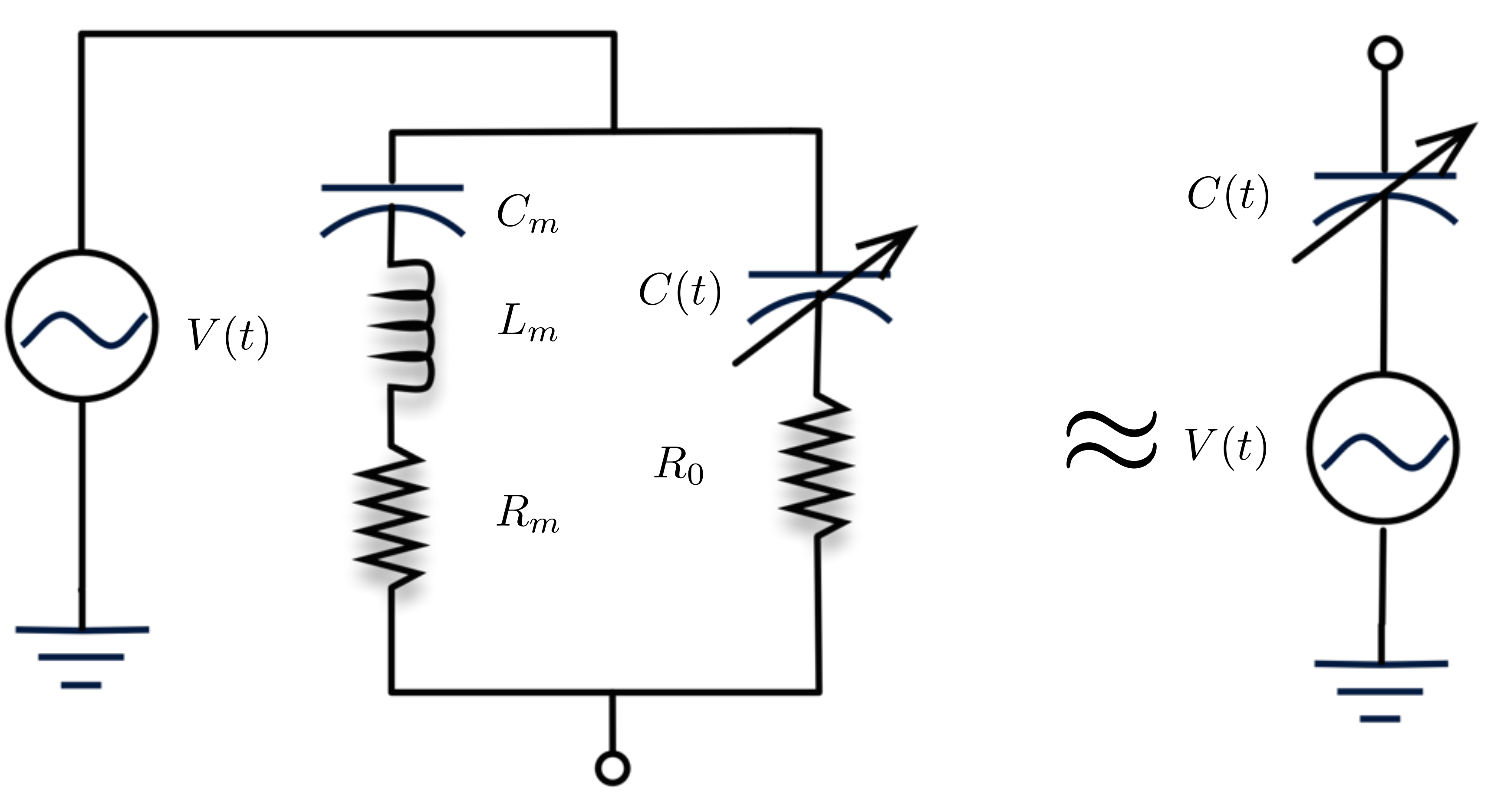}
  \caption{Equivalent circuit for the FBAR. We have employed the modified van Dyke-Butterworth model to provide the equivalent lumped-element circuit representation of the FBAR.} \label{fig:equiv_circuit}
\end{figure} 

Adding a series resistor $R_s$ at the input to this model allows to account for the electrode electrical loss. This is the so-called Butterworth-Van Dyke circuit, which has 5 parameters: $C_0,R_s,R_m,L_m,C_m$. One can add a sixth parameter, a resistor $R_0$ in series with $C_0$, accounting for material loss. In Ref.~\cite{LBWR00}, it is derived a set of equations to obtain the 6 parameters from measurable quantities: series and shunt resonant frequencies, the effective quality factors $Q_{s0},Q_{p0}$, and the capacitance and resistance far away from the resonances.

To model the \fbar{} we use experimental parameters closely following Ref.~\cite{OHAB10}: $C_m = 0.655$~fF, $L_m = 1.043$~$\mu$H and $R_m = 146$~Ohm for the mechanical part, and $C_0 = 0.4$~pF (average value in time) and $R_0 = 8$~Ohm for the geometric part. We would like to note that a similar \fbar{} was recently demonstrated with $C_0 = 1.00$~pF~\cite{RM12}. This higher coupling capacitance would allow increasing the DCE photon rate further. Taking into account that the permittivity of AlN is $\varepsilon_{\text{AlN}} \approx 9.2 \varepsilon_0 = 81$~pF/m and that the distance between the plates is $t_{\text{AlN}} = 350$~nm, we can model a parallel plate capacitor and estimate its area $A =t_{\text{AlN}} C_0 / \varepsilon \approx 7.7\cdot 10^{-10}~$m$^2$.

In this work, we are primarily interested in calculating the DCE photon production. Therefore, we simplify the full model to the elements containing the most relevant information, as shown in Fig.~\ref{fig:equiv_circuit}. Taking the parameters considered in \aref{app:params}, then $Z_m = R_m + j (\Omega L_m -\frac{1}{\Omega C_m}) \approx 146 + j 2.8 \cdot 10^4$~Ohm, while $Z_0 = R_0 -j \frac{1}{\Omega C_0} \approx 8 - j 189$~Ohm. As $Z_{\text{eq}} = \frac{Z_0 Z_m}{Z_0 + Z_m} \approx Z_0$, since $|Z_0|\ll |Z_m|$ and the resistor is much smaller than the capacitor, then our approximation follows.

In order to understand the connection between the mechanical properties described in the Appendix \ref{app:coupling} and the modified Butterworth-Van Dyke model, let us explain how the mechanical $Q$-factor which appears in Eq.~\eqref{amplitude} is described in terms of the electric elements of the circuit. This connection is deeply related, as one may expect, to the resistances in the circuit and the coupling of the losses of the circuit when coupled to other circuits. A detailed theoretical and experimental analysis of this dependence may be found in Ref. \cite{LBWR00}. In this reference, it is proven that 
\begin{equation}
    \frac{1}{Q} \approx \left (\frac{1}{Q_s}+\frac{1}{Q_e} \right) 
\end{equation}
with $(Q_s)^{-1}= \omega R_m C_m$, $(Q_e)^{-1} = \omega R_0 C_m$, and $\omega$ the resonance frequency of the circuit. Hence, the quality factor is indeed determined by the electric elements of the circuit. 

\subsection{Superconducting cavity}

The superconducting cavity has a length of $d=3.3 \cdot 10^{-2}$~m, with a fundamental frequency of $\omega_{c,0}=2\pi\cdot v/d=2\pi\cdot 3.03$~GHz with the speed of light in the superconducting material of $v=10^8$~m/s. It is capacitively coupled to a semi-infinite transmission line with impedance $Z_0 \approx 55$~Ohm. The capacitor coupling the resonator to the transmission line has a frequency $\omega_c/2\pi = (2 \pi Z_0 C_c)^{-1} \approx 2\pi\cdot 29.1$~GHz.\\

\subsection{Parameters for the proposed implementation}\label{app:params}

The following table summarizes the experimental parameters used in this proposal.

\begin{widetext}
\centering
\begin{tabular}{l l l}
	\hline\hline
	Parameter           & Symbol    & Value\\
	\hline
	\multicolumn{3}{l}{Material properties AlN}\\
	\hline
	Youngs modulus                  & $E$       &$308$\,GPa \\
	density                         & $\rho$    &$3230$\,kg/m$^3$ \\
	piezoelectric coupling \cite{lueng_piezoelectric_2000,dubois_properties_1999}          & $d_{33}$  &$5.1\cdot10^{-12}$\,m/V
	 \\
	Poissons ratio along \{0001\}   & $\nu$     &$0.287$  \\
	average velocity of sound       & $v$       &$9100$\,m/s \cite{OHAB10} \\
	permittivity                    & $\varepsilon_{\text{AlN}}$    &$9.2 \varepsilon_0$\\ 
	\hline\hline
	\multicolumn{3}{l}{FBAR parameters}\\
	\hline
	quality factor                  & $Q$       &$300$ \\
	resonance frequency             & $\Omega$  &$2\pi\cdot 4.2$~GHz \\
	thickness AlN layer             & $t_{\text{AlN}}$  &$350$\,nm \\
	drive voltage                   & $V_{\text{pp}}$   &$0.5$~mV\\
	driven motional amplitude       & $\Delta x(V_{\text{pp}})$  &$1.7\cdot V_{\text{pp}}$~nm/V\\
	\hline\hline
	\multicolumn{3}{l}{Superconducting cavity parameters}\\
	\hline
	length                          & $d$       &$3.3 \cdot 10^{-2}$~m\\
	cavity frequency    & $\omega_{c,0}$    &$2\pi\cdot3.03$~GHz\\
	impedance                       & $Z_0$     &$55$~Ohm\\
	DCE capacitance        & $C_0$              &$0.4\cdot10^{-12}$~F\\
	transmission line coupling rate      &  $\omega_{c}$ &$2\pi\cdot 29.1$~GHz\\
	\hline\hline
	\multicolumn{3}{l}{Further parameters}\\
	\hline
	Temperature                     & $T$    & $10$\,mK \\
	\hline\hline
\end{tabular}
\end{widetext}

\section{DCE miscellanea}

\subsection{DCE photon rate and its relation to $v/c$}\label{app:velocity}

In the following, we will relate the photon rate from \eref{eq:Nout} to the commonly used speed ratio $v/c$ in the DCE, where $v$ is the velocity of the mirror (in our case the velocity of the vibrational motion of the \fbar{}) and $c$ is the speed of light (in our case the speed in the superconducting material).

The mechanical photon production rate from \eref{eq:Nout} depends essentially on the term $\langle n_{\text{out}} (\omega)\rangle = | S_2^{\text{res}}(\omega,\Omega-\omega) |^2$, which is a function of $S(\omega',\omega'')  = -i \Delta C Z_0 \sqrt{|\omega'| |\omega''|} \theta(\omega') \theta(\omega'')$. Therefore, $\langle  n_{\text{out}} (\omega)\rangle \propto \Delta C^2 Z_0^{2}\omega(\Omega-\omega)$, which in resonance $\omega = \Omega/2$ gives $\langle  n_{\text{out}} (\Omega/2)\rangle \propto \Delta C^2 Z_0^{2}\Omega^2/4$. Taking into account that $\Delta C \approx C_0 \Delta x/t_{\mathrm{AlN}}$ and that the maximal speed of the FBAR is given by $v = \Delta x\cdot\Omega$, then $\langle  n_{\text{out}} (\Omega/2)\rangle \propto C_0^2 Z_0^{2}v^2/(4t_{\text{AlN}}^2)$. Finally, the speed of light in the material is given by $c = (\mathcal{C}Z_0)^{-1}$, with $\mathcal{C}$ the density of capacitance of the superconducting cavity, so 
\begin{equation}
\langle  n_{\text{out}} (\Omega/2)\rangle \propto C_0^2 \mathcal{C}^2 Z_0^{2}v^2/(4\mathcal{C}^2 t_{\text{AlN}}^2) = Q v^2/c^2,
\end{equation}
with $Q = C_0^2/(4\mathcal{C}^2 t_{\text{AlN}}^2)$ is related with the impedance mismatch between the FBAR and the superconducting cavity and, consequently, with the quality factor of the resonator. This expression coincides with Eq.~(15) of Ref.~\cite{LJR96} and Eq.~(3) of Ref.~\cite{JJWN10} for photon production in the presence of a cavity, and shows that, effectively, in the context of the FBAR scheme, the $v/c$ ratio in photon production still holds.

For the sake of completeness, let us estimate the ratio of $v/c$ as follows. The maximal velocity of the mechanical resonator on resonance is given as $v=\Delta x\cdot\Omega$. The speed of light in the superconducting cavity is $c=(\mathcal{C}Z_0)^{-1}=1\cdot 10^8$\,m/s. We obtain with $\Omega=2\pi\cdot 4.2$GHz a ratio $v/c$ of $\sim 2\cdot 10^{-10}$ and $\sim 2\cdot 10^{-8}$ for $\Delta x=8.5\cdot 10^{-13}$\,m (low mechanical Q) and $\Delta x=8.5\cdot 10^{-11}$\,m (high mechanical Q), respectively.

\subsection{DCE as a parametric effect}\label{app:parametric}

In the following, we show by using a simplified model how the parametric effect of the DCE emerges from a change in boundary conditions. To this end, we use the Hamiltonian of a cavity ended by an oscillating mirror, with $C_T = C + C_0$, the sum of the total capacitance of the cavity and the capacitor. Therefore,
\begin{equation} \label{Hamiltonian}
H = \frac{1}{2 C_T} q^2 + \frac{1}{2L} \Psi^2,
\end{equation}
which is the Hamiltonian corresponding to an $LC$ circuit. Let us assume that the face of the resonator vibrates due to the phonons as a classical harmonic oscillator of frequency $\Omega$ and amplitude $\Delta x$, so $d(t) = d_0 + \Delta x \cos (\Omega t)$ and the capacitance, that we assume given by a coplanar capacitor, varies as
\begin{align}\label{capt}
\frac{1}{C_T (t)} & = \frac{1}{C + C_0(t)} = \frac{1}{C + \frac{\epsilon A}{d(t)}} \nonumber\\ 
& = \frac{1}{C+C_0(1-\frac{\Delta x}{d_0} \cos (\Omega t))} \nonumber\\
& =\frac{1}{C_T}+ \frac{C_0 \Delta x}{C_T^2 d_0} \cos (\Omega t)
\end{align}
We can consider the term on the right hand side a perturbation, so the Hamiltonian can be expressed by means of the same creation and annihilation operators as
\begin{equation*}
H = \hbar \omega \left (a^{\dagger} a + \frac{1}{2} \right ) - \frac{C_0 \Delta x}{2 C_T^2 d_0} \frac{\hbar \omega C_T}{2} \cos (\Omega t) \left (a^{\dagger} -a \right)^2
\end{equation*}
where we made use of the standard definition of creation and annihilation operators for an $LC$ circuit $q = i \sqrt{\frac{\hbar \omega C_T}{2}} (a^{\dagger} - a)$ and $\Psi = \sqrt{\frac{\hbar}{2 \omega C_T}} (a^{\dagger} + a)$, and $\omega = \sqrt{\frac{1}{LC_T}}$ is the frequency of the superconducting circuit. By removing constants and using the rotating wave approximation with $\Omega = 2 \omega$, one can rewrite the Hamiltonian in the interaction picture as
\begin{equation}\label{Hamil}
H = \frac{\hbar \omega}{8}\frac{C_0 \Delta x}{C_T d_0}  \left [(a^{\dagger})^2 + a^2 \right ],
\end{equation}
which is a squeezing Hamiltonian.

\end{document}